\documentclass[ aps, 12pt, final, notitlepage,
twoside,onecolumn, nobibnotes, nofootinbib,
superscriptaddress,noshowpacs, centertags] {revtex4}
\begin{document}
\selectlanguage{english}

\title{A Search for Periodicity in the Light Curves of Selected Blazars}

\author{\firstname{N.~A.}~\surname{Kudryavtseva}}
\email[]{nadia@gong.astro.spbu.ru} \affiliation{Astronomy
Institute, St. Petersburg State University, St. Petersburg,
Russia}

\author{\firstname{T.~B.}~\surname{Pyatunina$^\dag$}}
\affiliation{Institute of Applied Astronomy, St. Petersburg,
Russia}

\date{\today}

\begin{abstract}
We present an analysis of multifrequency light curves of the
sources $2223-052$ (3C~446), $2230+114$ (CTA~102), and $2251+158$
(3C~454.3), which had shown evidence of quasi-periodic activity.
The analysis made use of data from the University of Michican
Radio Astronomy Observatory (USA) at 4.8, 8, and 14.5 GHz, as well
as the Metsahovi Radio Astronomy Observatory (Finland) at 22 and
37 GHz. Application of two different methods (the discrete
autocorrelation function and the method of Jurkevich) both
revealed evidence for periodicity in the flux variations of these
sources at essentially all frequencies. The periods derived for at
least two of the sources -- $2223-052$ and $2251+158$-- are in
good agreement with the time interval between the appearance of
successive VLBI components. The derived periods for $2251+158$ ($P
= 12.4$ yr and $2223-052$ ($P = 5.8$ yr) coincide with the periods
found earlier by other authors based on optical light curves.
\end{abstract}

\maketitle

\section{INTRODUCTION}
Studies of the variability of active galactic nuclei (AGNs)
represent an effective tool for investigating the nature of the
activity and the mechanisms leading to its observable
manifestations. The long-term variability of AGNs, including
possible periodicity of the variations, has been studied and
discussed in detail in a number of
works~\cite{aller2003,Kelly2003,Ciaramella2004,Fan2002}.
Traditionally, a large fraction of studies dedicated to searches
for periodicity have been based on optical observations, since the
durations of the historical light curves in the optical are as
long as $\sim100$~yr for some sources, while the maximum durations
of radio light curves are no more than $\sim35$~yr. Although the
results of such studies have often been contradictory, the
detection in 1995 of an optical flare in OJ~287 that had been
predicted based on the period of 11.9 yr detected in the
historical light curve provided convincing confirmation of the
presence of periodic activity in this source~\cite{Pursimo2000}.
Possible periodicity in the radio activity has also recently been
detected for several
objects~\cite{Kelly2003,Ciaramella2004,Pyatunina2005}. It is
interesting that the radio period found for AO 0235+164 coincides
with the optical period~\cite{Raiteri2001}. This coincidence of
the optical and radio periods testifies that the radiation in both
wavebands has the same (synchrotron) nature, and arises in
coincident or nearby regions. A similar conclusion was drawn
in~\cite{Lister2005} based on a comparison of the properties of
the polarized emission in the optical and in the radio at 43GHz.

As we indicated above, data on periodicity of the activity of AGNs
are often contradictory or inconsistent. In some cases, different
periods have been derived for a single object based on analyses of
data for different time intervals, which could have both
methodical and physical origins. From a methodical point of view,
the lack of agreement between derived periods could be a
consequence of the limited extent and irregularity of the
observational series used, which leads to a period found using one
section of the light curve differing from that found using a
different section. In addition, the physical processes giving rise
to quasi-periodic variations of the flux density can be divided
into two types: primary processes that are associated with the
periodic generation of perturbations at the "central engine" due
to multiplicity of the central black hole, and secondary processes
that arise during the propagation of perturbations along the jet,
such as the propagation of shocks along the
jet~\cite{gomez1997,Pyatunina2000} or
recollimation~\cite{gomez2005} or precession of the
jet~\cite{Stirling2003}. It is obvious that secondary processes
are more likely to be manifest on short time intervals, but, the
longer the duration of a series of observations, the more likely
it is that periodicity due to primary processes will appear.
Quasi-periodic processes with various natures forming in various
regions of the jet can also be superposed, complicating and
confusing the pattern of the periodicity.

Currently, the most widely accepted hypothesis explaining both
periodicity in the activity of AGNs and the activity itself is the
presence of a multiple black hole at the galactic
nucleus~\cite{lobanov2002}. One drawback of this theory is the
short lifetime of a system of black holes due to the loss of
energy to gravitational radiation. This problem can be avoided if
we consider the multiplicity of the black holes as a dynamical
process, where the main black hole captures new companions on time
scales that are comparable to the lifetime of the previous
black-hole pair~\cite{Haehnelt1994,Kidger2000}.

As was shown in~\cite{Sundelius1997}, in models with multiple
black holes, the process leading to the appearance of the primary
perturbations in the vicinity of the multiple black hole may not
have a strictly periodic character. Therefore, we will think in
terms of the presence of quasiperiods in the light curves of AGNs,
rather than some strict periodic process. A good example is
provided by OJ~287, which is the best observed and studied object
in this context, and shows the presence of precisely quasiperiodic
flux-density variations~\cite{Kidger2000}. Nevertheless, the
presence of such quasiperiodic behavior can be used to within
known limits predict the epochs of flares, making it possible to
verify multiple black-hole models, study the propagation of the
associated perturbations along the jet in more detail, and plan
future observations based on the expected activity phase of the
source.

The monitoring of AGNs carried out at 4.8, 8, and 14.5 GHz at the
University of Michigan Radio Astronomy Observatory (UMRAO,
USA)~\cite{Aller1996} and at 22 and 37 GHz at the Metsahovi Radio
Astronomy Observatory (MRAO, Finland)~\cite{Terasranta1998} over
more than 30 yr present unique opportunities for studies of the
long-term evolution of these objects. These databases have been
analyzed earlier on the basis of structure
functions~\cite{Hughes1992,Lainela1993}, wavelet
analysis~\cite{Kelly2003}, and autocorrelation
matrices~\cite{Ciaramella2004}, but the information that can be
extracted for these data has not been exhausted.

In our earlier works~\cite{Pyatunina2005,Pyatunina2004}, based on
an analysis of multifrequency light curves of selected blazars, we
divided the observed radio flares into core events, which display
low-frequency delays and are associated with primary perturbations
in the core, and jet events, which appear simultaneously at all
frequencies and are associated with the propagation of primary
perturbations along the jet~\cite{Pyatunina2000,Zhou2000}.We also
defined the time intervals between successive core flares as the
characteristic time scale for the development of activity in these
blazars. The questions of how the characteristic time scale for
the development of activity varies from source to source and
whether these time scales are preserved over long times within a
single source are of fundamental importance for our understanding
of the nature of the activity. We present here estimates of
possible periodic components in the light curves of several
sources we have studied earlier, derived using the method of
Jurkevich and the discrete autocorrelation function. The advantage
of these methods is that they are independent of the form of the
signal, and can be used to find periodicities not only when the
signal is sinusoidal, as in the case of Fourier analysis.

\section{LIGHT CURVES}
We used the UMRAO and MRAO monitoring databases to carry out our
searches for periodicity. Based on the results of our previous
detailed analyses of flares in several
blazars~\cite{Pyatunina2005,Pyatunina2004}, we selected three
objects showing evidence for possible quasiperiodic flux
variations. Table~\ref{table1} presents a list of these objects,
the observing frequencies $\nu$ in GHz, the range of dates for
which observations are available, the duration of the
observational series in years $\Delta T$, and the number of flux
measurements $N$. The durations of the observational series are
very different for different objects and for different
frequencies, and ranges from 17 to 38 yr.

\section{METHOD FOR ANALYSIS OF THE LIGHT CURVES}

\subsection{Discrete Autocorrelation Function}
The application of the discrete autocorrelation
function~\cite{Edelson1988} is designed to calculate the
correlation coefficient between two series of irregularly
distributed data as a function of the time shift between them. The
method can also be used to search for periodicity if the two data
series are identical. If there is a period in the light curve, the
discrete autocorrelation function should exhibit a clear
correlation at time shifts equal to zero and to the period. The
closer the correlation coefficient is to unity, the more
trustworthy is the identified period.

To determine the values of the discrete autocorrelation function
for each pair of data $(a_{i}, b_{j})$, we calculate the function
$UDCF$:
\[ UDCF_{ij} = \frac{(a_{i} - \bar{a})(b_{j} - \bar{b})
}{\sqrt{\sigma_a^2 \sigma_b^2 }}, \]

where $\bar{a}$, $\bar{b}$ are the mean values of the data series
and $\sigma_{a}$, $\sigma_{b}$ are the corresponding standard
deviations. Each pair of data is associated with a time interval
$\Delta t_{ij} = t_{j} - t_{i}$. The set of values $UDCF_{ij}$ are
divided into groups so that the time interval between the
observation times $\Delta t_{ij}$ falls into some interval around
the trial time delay: $\tau - \Delta \tau / 2 \leq \Delta t_{ij}
\leq \tau + \Delta \tau / 2$. Averaging the $UDCF_{ij}$ values
over each group yields the discrete autocorrelation function:
\[ DCF(\tau) = \frac{1}{M} \sum UDCF_{ij}(\tau), \]
where $M$ is the number of pairs in the group. The error is
calculated using the formula
\[ \sigma(\tau) = \frac{1}{M - 1} \{ \sum [UDCF_{ij} - DCF(\tau) ] ^2
\}^{1/2}. \] The discrete autocorrelation function often displays
a flat peak. In such situations, we approximated this function
with a Gaussian in order to more accurately estimate the period.

\subsection{Method of Jurkevich}
\subsubsection{Description of the method.}
The method of Jurkevich~\cite{Jurkevich1971} is based on analysis
of the dispersions of the phase curves constructed for a series of
trial periods. The essence of the method is that, the closer a
trial period to the real period, the smaller the scatter of the
points in the phase curve. To identify the period for which the
dispersion reaches a minimum, phase curves are constructed for a
series of trial periods, where the phase is defined by the
relation
\[ \varphi_{i} = \frac{T - T_{0}}{P} \; \; (mod \; 1), \]
where $P$ is the trial period and $T_{0}$ is the time
corresponding to the zero phase. The time $T_{0}$ is usually taken
to be the middle of the observational interval. Then, all the data
are divided into $m$ groups according to their phases, and the rms
deviation is calculated for each group:
\[ V_{l}^2 = \sum_{i=1}^{m_{l}} x_{i}^2 - m_{l} \bar{x_{l}}^2,\]
where $m_{l}$ is the number of points in the group and
$\bar{x_{l}}$ is the mean value $x_{i}$ for the group. Further,
the sum of the mean square deviations over all groups is
calculated:
\[ V_{m}^2 = \sum_{i=1}^{m} V_{l}^2. \]
The appearance of a minimum in the function $V_{m}^2$ at some
trial period could indicate that this corresponds to a real period
in the variations. Kidger et al.~\cite{Kidger1992} proposed the
use of a modified function $V_{m}^2$ to estimate the
trustworthiness of the identified periods:
\[ f = \frac{1 - V_{m}^2}{V_{m}^2}, \]
where here $V_{m}^2$ is normalized to the value $V_{l}^2$
calculated for the entire dataset. A value $V_{m}^2 = 1$
corresponds to $f = 0$, and denotes an absence of periodicity in
the data series. The values of periods can be found from a plot of
the function $f$: values $f  \geq 0.5$ indicate very strong
periodicity, while values $f  \leq 0.25$ indicate that, if any
periodicity is present, it is manifest only very weakly. Another
criterion for the trustworthiness of a period is the depth of the
minimum of $V_{m}^2$ relative to the noise. If this minimum is a
factor of ten larger than the error in the "flat" part of the
curve, the period corresponding to this minimum can be considered
real~\cite{Kidger1992}. We fit Gaussians to the plots of the
modified function $V_{m}^2$ in order to more accurately find the
position of the minimum. We estimated the error in the resulting
period formally as the half-width of the corresponding peak.

\subsubsection{Search for false periods using Monte-
Carlo simulations.} The irregularity of the observational series
can lead to the appearance of false periods. We carried out
Monte-Carlo simulations in order to separate out periods
associated with real variability of the objects from those
associated with the irregularity of the observations. The
synthetic light curves generated in this way were analyzed using
the method of Jurkevich. The artificial light curves were
constructed from the real observations using the method presented
in~\cite{Fan2002}: the observation times were retained, and the
fluxes were chosen at random from the real light curve. Thus, any
period found for these light curves can only be a consequence of
the time distribution of the observations.

If periods detected in the synthetic data coincided with periods
found for the real light curves, they were classified as false and
were not considered further.

\section{RESULTS OF THE ANALYSIS}
\subsection{2223-052 (3C~446)}
The source $2223-052$ (z = 1.404) is known to exhibit
characteristics of a quasar or a BL Lac object, depend on its
activity state. In its quiescent state, it displays strong
emission lines characteristic of quasars, while a powerful
continuum dominates in the active phase, giving rise to weak
emission lines, as is characteristic for BL Lac
objects~\cite{barbieri1985}. Periods of 4.2 and 5.8 yr have been
detected in the optical~\cite{barbieri1985}; the period of 4.2 yr
was subsequently confirmed in multiple
studies~\cite{webb1988,barbieri1990}.

\begin{figure}
\setcaptionmargin{3mm} \onelinecaptionsfalse
\includegraphics{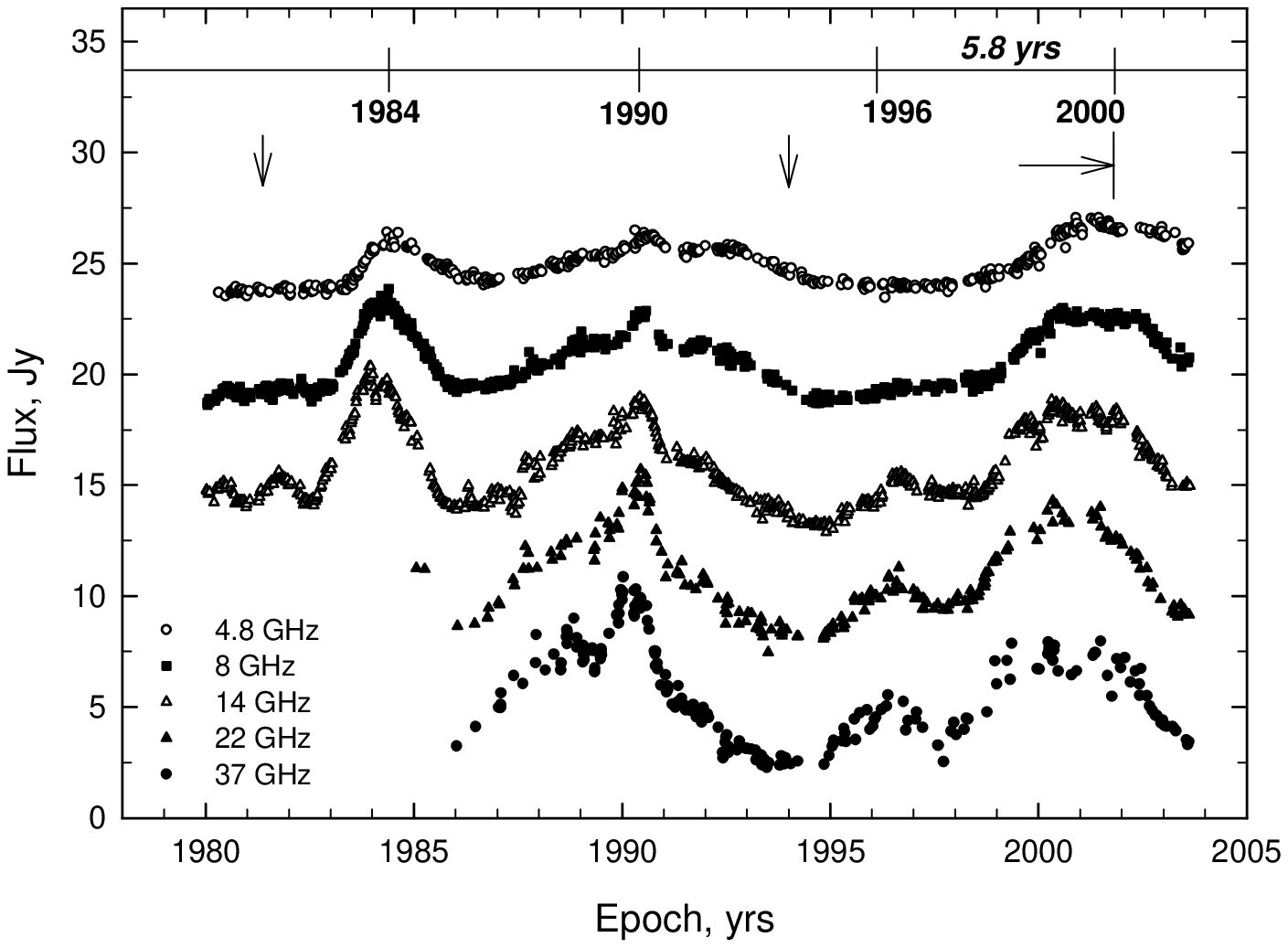}
\captionstyle{normal} \caption{ Light curves of 2223-052 for 37,
22, 14.5, 8, and 4.8 GHz. For ease of viewing, the curves have
been shifted relative to each other by 5 Jy. The arrows mark the
birth epochs of VLBI components.} \label{2223-lightcurve}
\end{figure}

\begin{figure}
\setcaptionmargin{5mm}
\onelinecaptionsfalse
\includegraphics{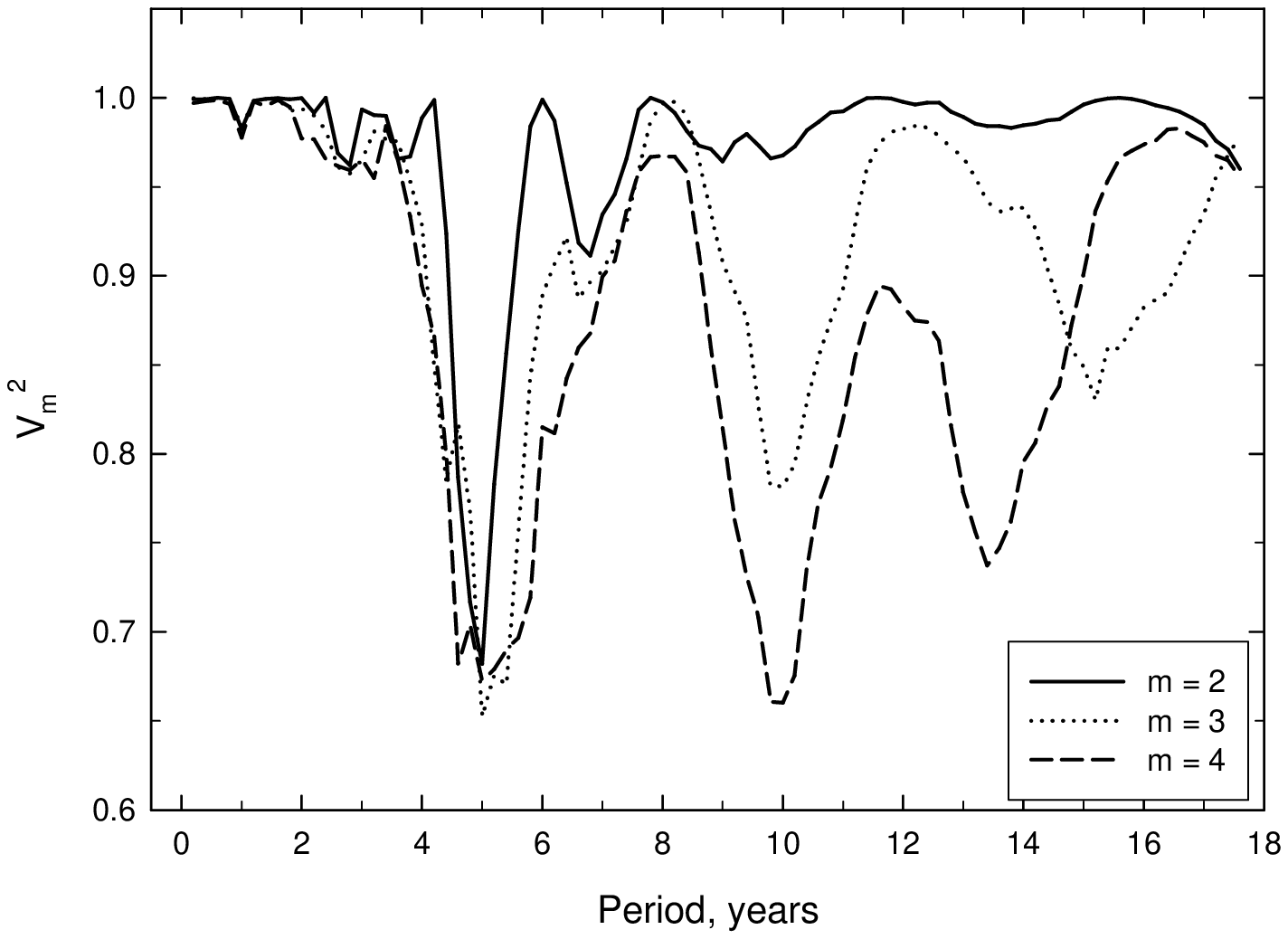}
\captionstyle{normal} \caption{Results of the period search for
2223-052 at 8 GHz using the method of Jurkevich. Plots for various
bin numbers for the phase curve are presented.} \label{2223-jur-8}
\end{figure}

Figure~\ref{2223-lightcurve} presents the light curves from 4.8 to
37 GHz. Flares of two types alternate in the light curve: narrow,
single flares with inverted spectra (growing toward higher
frequencies), and broad flares with nearly flat spectra. The time
interval between a narrow flare and the following broad flare is
about 6 yr, while the interval between two successive flares of
the same type is about 12 yr. Applying the method of Jurkevich to
the light curves of $2223-052$ yielded with a high degree of
confidence the presence of periods of $5.8 \pm 0.5$ yr for all
five frequencies, and $9.8 \pm 0.5$ yr for all frequencies except
4.8 GHz (here and below, the presented periods are the average
values over the various frequencies). Figure~\ref{2223-jur-8}
presents the plot of the function $V_{m}^2$ for the 8 GHz data,
which illustrates the results. It is interesting that the detected
period $5.8 \pm 0.5$ yr coincides with the period of 5.8 yr found
in~\cite{barbieri1985} for the optical light curve of $2223-052$.
However, further observations in both the radio and optical are
needed in order to confirm this agreement of the periods of
$2223-052$ in these two wavebands.

\begin{figure}
\setcaptionmargin{2mm} \onelinecaptionsfalse
\includegraphics{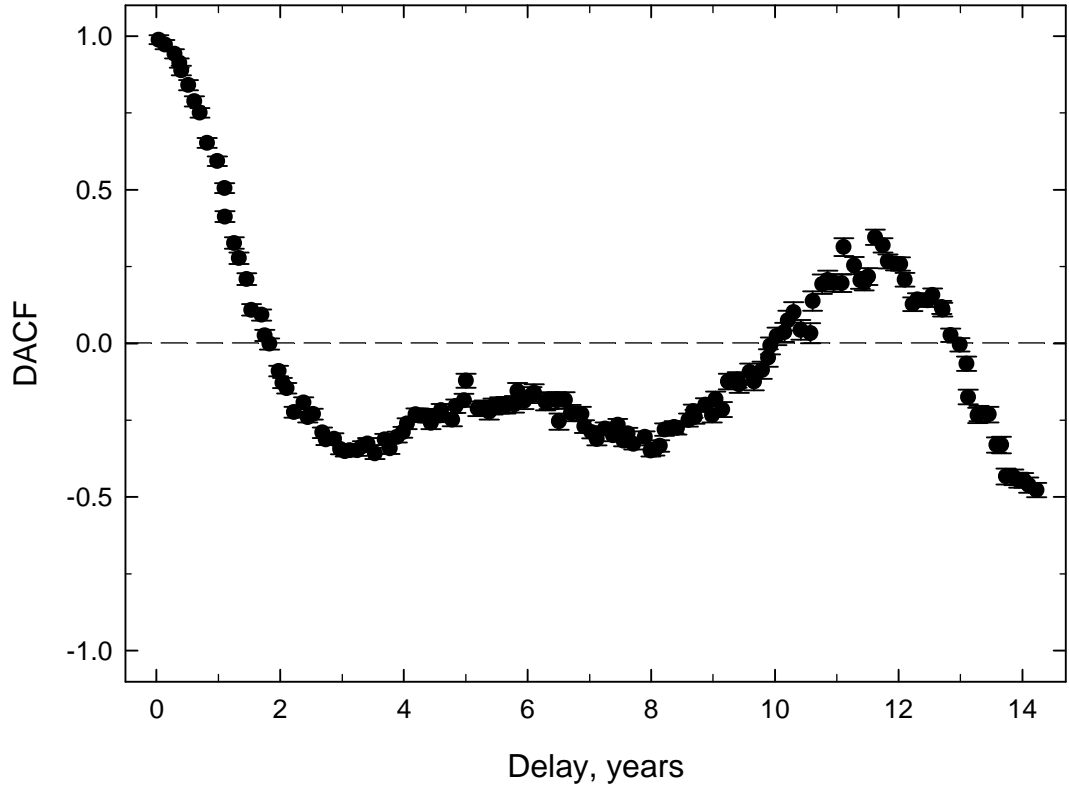}
\captionstyle{normal} \caption{Plot of the discrete
autocorrelation function constructed for 2223-052 at 14.5 GHz.}
\label{2223-dcf-14}
\end{figure}

The discrete autocorrelation function yielded a period of $10.9
\pm 0.2$ yr for 37, 22, and 14.5 GHz. Figure~\ref{2223-dcf-14}
presents the discrete autocorrelation function for 14.5 GHz.
Table~\ref{table2223} gives a list of the derived periods, where
$\nu$ is the observing frequency in GHz, $P_{Jurk}$ the period
found using the method of Jurkevich (in years), $f$ the confidence
measure for $P_{Jurk}$, $P_{DACF}$ the period found using the
discrete autocorrelation function (in years), and $k$ the
corresponding correlation coefficient.

Figure~\ref{2223-lightcurve} presents light curves for $2223-052$
for all the frequencies, with the flares corresponding to the
period of $5.8 \pm 0.5$ yr found with a high confidence level f
using the method of Jurkevich indicated. We can see that is period
describes the activity of the source well, i.e., the flares of
1984, 1990, 1996, and 2000. On the other hand, the discrete
autocorrelation function detected only the 11-year period,
possibly due to the coincidence in the structure of the "broad"
and "narrow" flares when the light curve is shifted by 11 years.
The question of both types of flares are associated with the birth
of new superluminal components can be addressed using VLBI
observations at 22 or 43 GHz .

The blazar $2223-052$ was observed by Kellermann et
al.~\cite{kellerman2004} as part of a program of VLBI monitoring
of AGNs at 15 GHz. These 15 GHz observations can provide
information about large-scale processes in the evolution of the
source and indicate the characteristic time scale for activity
cycles. Kellermann et al.~\cite{kellerman2004} report the births
of VLBI components at epochs $1981.4 \pm 1.6$ and $1994.0 \pm
0.8$, marked on the light curve (Fig.~\ref{2223-lightcurve}) by
vertical arrows. Given the possible uncertainties in determining
the birth epochs of the components, these epochs are fairly close
to the powerful flares of 1984 and 1996. A visual analysis of the
maps obtained in the MOJAVE 15-GHz monitoring
program~\cite{Lister2005}, available on the MOJAVE website,
suggest the birth of a new VLBI component in 2002. We can see from
Fig.~\ref{2223-lightcurve} that this component could be associated
with the decay of a powerful flare in 2000. Taking into account
the uncertainties in the birth epochs of VLBI components and the
absence of VLBI observations before 1995, the births of observed
components agree fairly well with the epochs of the powerful
flares described by the period of $5.8 \pm 0.5$ yr.

Note that only four six-year periods, and slightly more than two
11-year periods, are present in the light curve of $2223-052$
(Table~\ref{table1}). Therefore, further observations at all
frequencies are required in order to verify these periods.
According to our estimates, the following flare should occur in
2007, and should be accompanied by the birth of a new component on
VLBI scales. Taking into account the redshift of the source, $z =
1.404$, and applying the relation $P = P_{obs}/(1+z)$, the
detected periods in the rest frame of the source are 2.4 and 4.5
yr.

\subsection{2230+114 (CTA~102)}
$2230+114$ is a gamma-ray blazar at $z = 1.037$~\cite{blom1995}.
In 1965, Sholomitsky~\cite{Sholomitsky1965} published the first
report of variability of this source, with a period of 102 days at
902 MHz. Although this result was not subsequently
confirmed~\cite{Terzian1966}, variability was soon detected at
shorter wavelengths~\cite{Pyatunina2005}.

\begin{figure}
\setcaptionmargin{5mm}
\onelinecaptionsfalse
\includegraphics{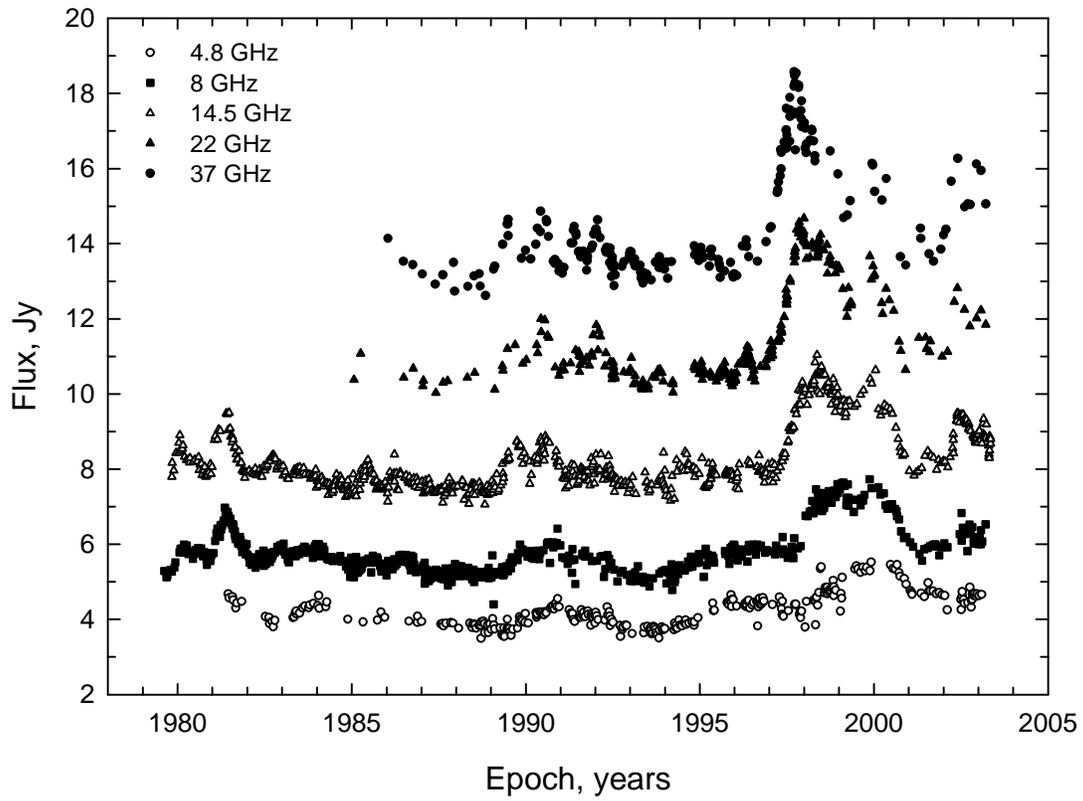}
\captionstyle{normal} \caption{Light curves of 2230+114 for 37,
22, 14.5, 8, and 4.8 GHz. For ease of viewing, the curves have
been shifted relative to each other in flux.} \label{2230-lcurve}
\end{figure}

Figure~\ref{2230-lcurve} presents the light curves of $2230+114$
at all five frequencies. We can see that bright flares are
observed approximately every eight years. The flares are
characterized by the presence of fine structure on time scales
appreciably shorter than one year. The amplitudes of the flares
vary appreciably, but the positions of individual features in the
flare structure are preserved~\cite{Pyatunina2005}. It is
interesting that not only the total amplitude of the flares in
each eight-year cycle, but also the relative brightnesses in the
fine-structure features, vary with time. In the first and third
eight year cycles, the first subflare dominates, while the two
first subflares have equal amplitudes in second cycle. All this
gives rise to a complex pattern for the variability, complicating
a mathematical analysis.

\begin{figure}
\setcaptionmargin{5mm}
\onelinecaptionsfalse
\includegraphics{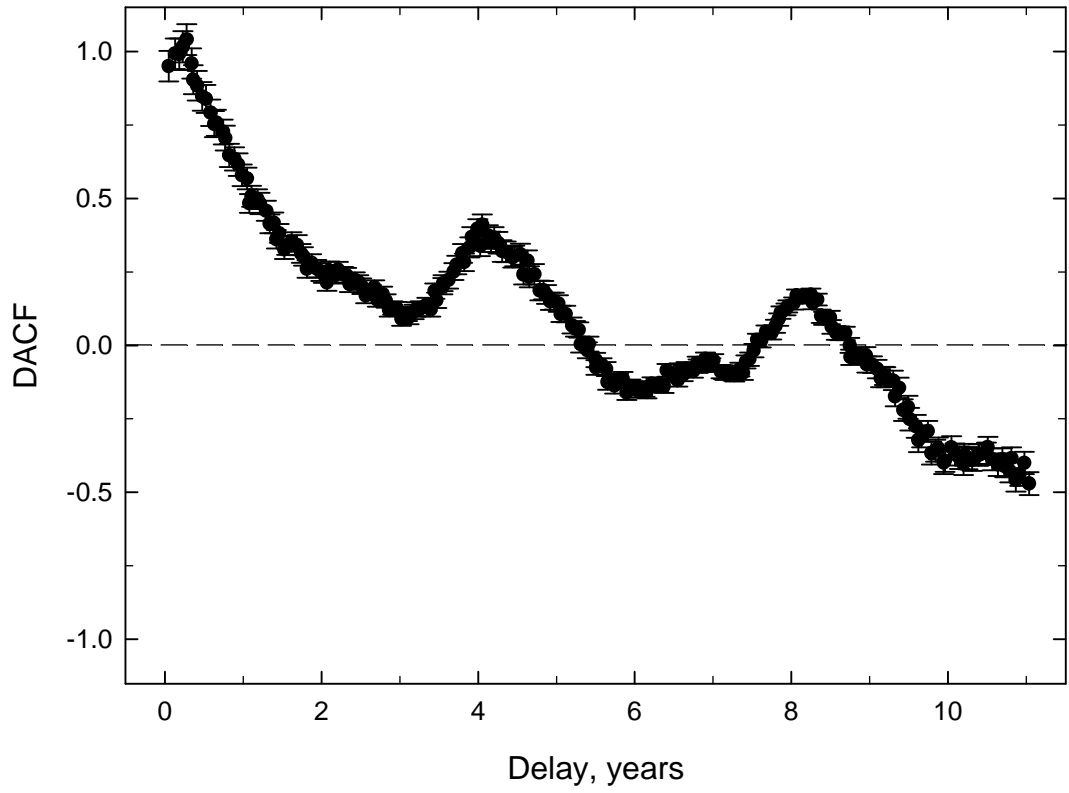}
\captionstyle{normal} \caption{Discrete autocorrelation function
for 2230+114 at 14.5 GHz.} \label{2230-dcf-14}
\end{figure}

\begin{figure}
\setcaptionmargin{5mm}
\onelinecaptionsfalse
\includegraphics{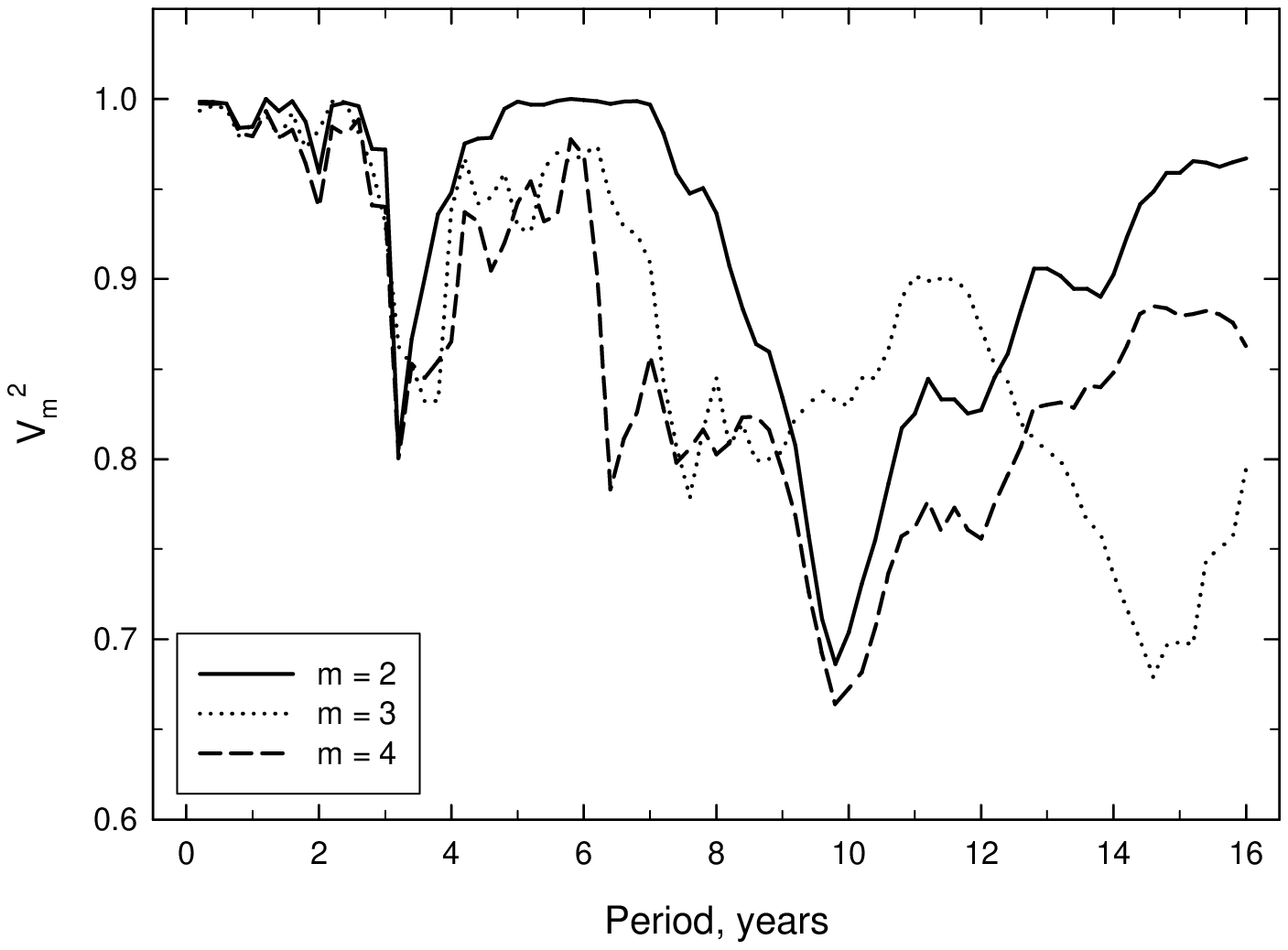}
\captionstyle{normal} \caption{Results of the period search for
2230+114 at 4.8 GHz using the method of Jurkevich. Plots for
various bin numbers for the phase curve are presented.}
\label{2230-jur-4}
\end{figure}

\begin{figure}
\setcaptionmargin{5mm}
\onelinecaptionsfalse
\includegraphics{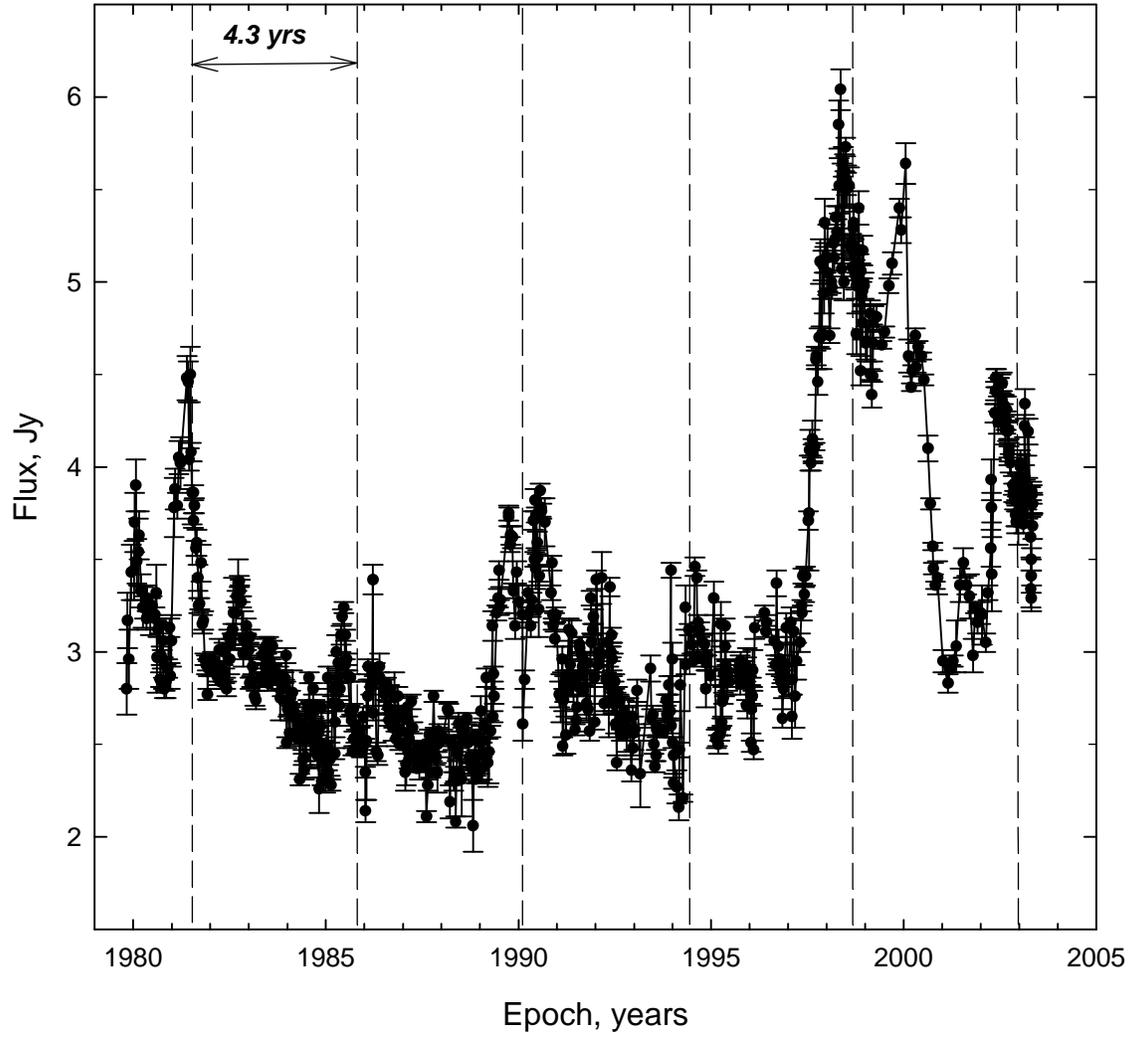}
\captionstyle{normal} \caption{7. Light curve of 2230+114 at 14.5
++¡. The dotted lines mark intervals corresponding to the $4.3 \pm
0.5$-year period.} \label{2230-14per}
\end{figure}

The discrete autocorrelation function yielded periods of $4.3 \pm
0.5$ yr at 14.5 GHz and $8.4 \pm 0.1$ yr at 14.5 and 4.8 GHz
(Table~\ref{table2230}). Figure~\ref{2230-dcf-14} presents the
discrete autocorrelation function for 14.5 GHz. The method of
Jurkevich yields periods of $4.6 \pm 0.7$ yr at 37, 22, and 4.8
GHz and $9.3 \pm 0.6$ yr at all frequencies except 8 GHz
(Table~\ref{table2230}, Fig.~\ref{2230-jur-4}).
Figure~\ref{2230-14per} presents the 14.5 GHz light curve with the
intervals corresponding to the period $4.3 \pm 0.5$ yr indicated.
We can see that this period describes well both the powerful
flares of 1981, 1990, and 1999 and the weaker flares 1986, 1995,
and 2003, while the period $8.4 \pm 0.1$ yr describes only the
large-amplitude flares (1981, 1990, and 1999).

\begin{figure}
\setcaptionmargin{5mm}
\onelinecaptionsfalse
\includegraphics{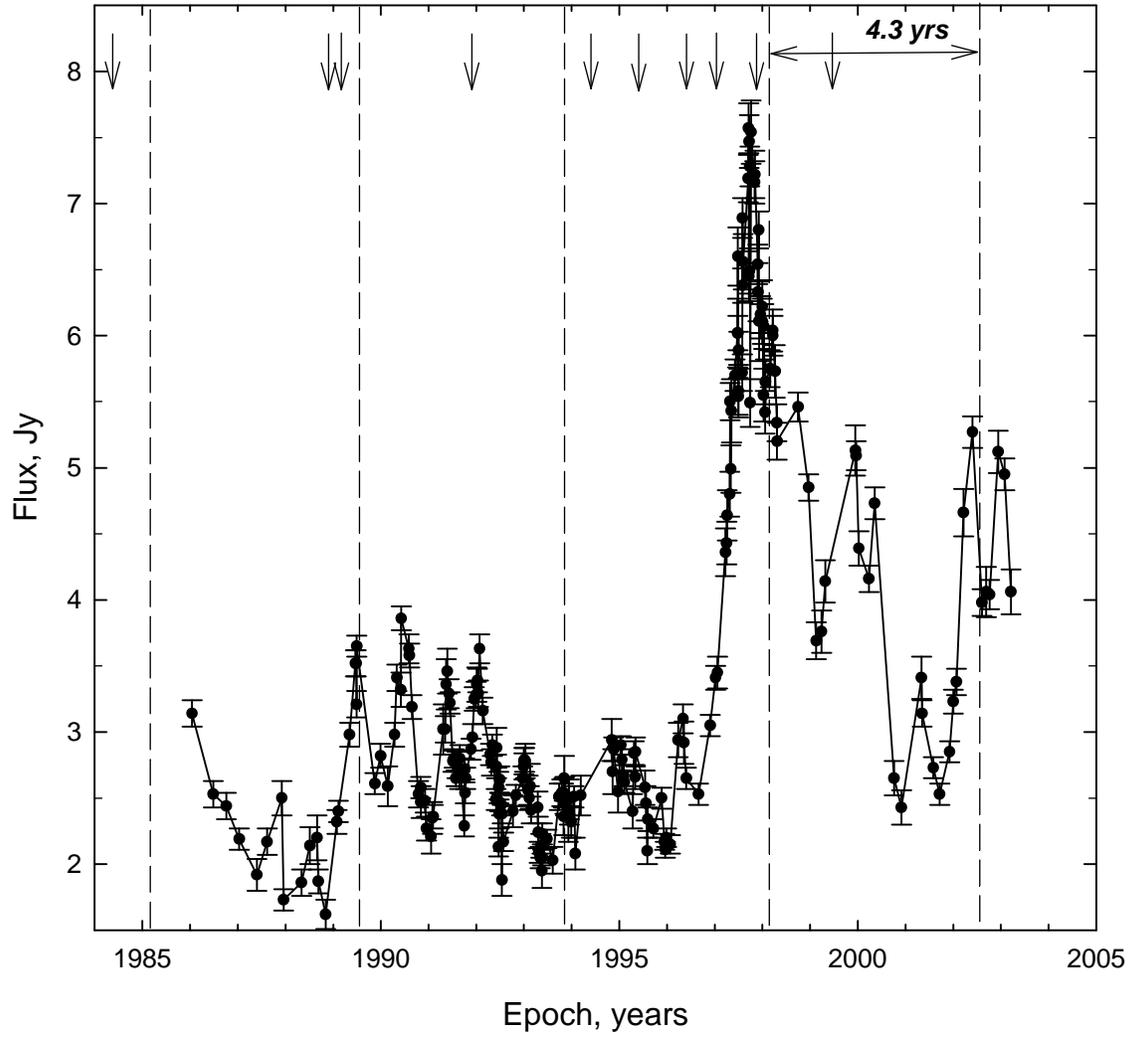}
\captionstyle{normal} \caption{Light curve of 2230+114 at 37 GHz.
The dotted lines mark intervals corresponding to the $4.3 \pm
0.5$-year period. The arrows mark the birth epochs of VLBI
components at
43GHz~\protect\cite{Jorstad2001,Jorstad2005,Rantakyro2003}.}
\label{2230-37per}
\end{figure}

Jorstad et al.~\cite{Jorstad2001,Jorstad2005} reported the birth
of VLBI components in $2230+114$ at epochs $1994.28 \pm 0.02$,
$1995.19 \pm 0.04$, $1996.08 \pm 0.02$, $1997.9 \pm 0.2$, and
$1999.54 \pm 0.04$ at 43GHz . Rantakyr\"{o} o et
al.~\cite{Rantakyro2003} estimated the birth epochs and speeds of
jet components based on all available VLBI observations in the
literature. The birth epochs of the 43-GHz VLBI components are
marked by arrows on the 37 GHz light curve
(Fig.~\ref{2230-37per}). Each birth epoch coincides with a flare
at 37 GHz, but the irregular and insufficiently dense series of
VLBI observations prevents a detailed comparison of the evolution
of the light curve with the evolution of the VLBI source
structure. The detailed light-curve analysis of Pyatunina et
al.~\cite{Pyatunina2005,Pyatunina2004} showed the presence of
activity cycles with a characteristic variability time scale of
$8.04 \pm 0.3$ yr, in good agreement with the period we have found
here, $8.4 \pm 0.1$ yr.

The available light curve encompasses about six four-year and
three eight-year periods (Table~\ref{table1}), and further
observations are required to verify these periods. According to
our estimates, the following flare should occur in 2007. Taking
into account the source redshift, the derived periods in the rest
frame of the source are 2.1 and 4.1 yr.

\subsection{2251+158 (3C~454.3)}
$2251+158$ is a gamma-ray blazar with a redshift of $z =
0.859$~\cite{blom1995}. By identifying the main components of the
autocorrelation matrix using a neural network, Ciaramella et
al.~\cite{Ciaramella2004}  found the period 6.07--6.55 yr for this
object at 4.8, 8, 14.5, 22, and 37 GHz. Cheng-yue~\cite{Su2001}
found a period of 12.39 yr in the optical based on a hundred-year
historical light curve in the $B$ band.

\begin{figure}
\setcaptionmargin{5mm}
\onelinecaptionsfalse
\includegraphics{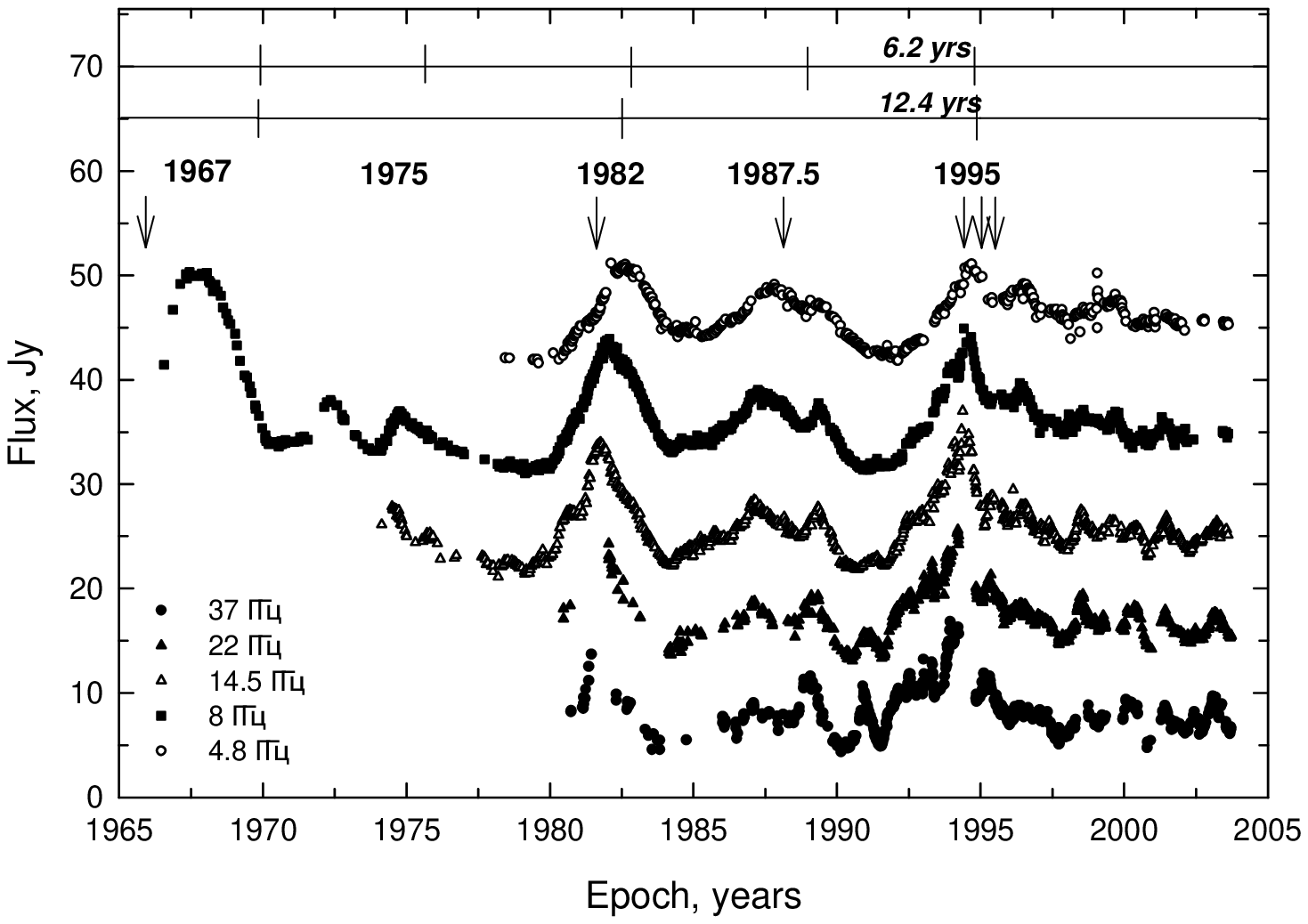}
\captionstyle{normal} \caption{Light curves for 2251+158 at 37,
22, 14.5, 8, and 4.8 GHz. For ease of viewing, the curves have
been shifted relative to each other by 8 Jy. The arrows mark the
birth epochs of VLBI components.} \label{2251-lcurve}
\end{figure}

\begin{figure}
\setcaptionmargin{5mm}
\onelinecaptionsfalse
\includegraphics{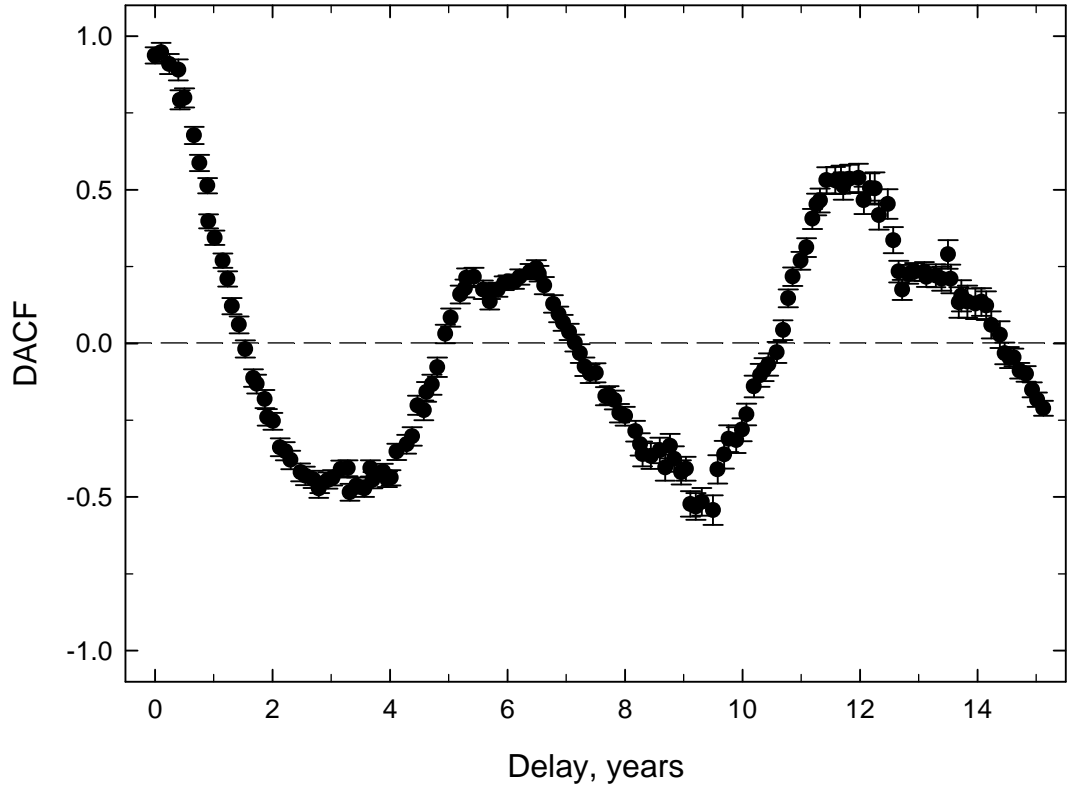}
\captionstyle{normal} \caption{Discrete autocorrelation function
for 2251+158 at 4.8 GHz.} \label{2251-dcf-4}
\end{figure}

\begin{figure}
\setcaptionmargin{5mm}
\onelinecaptionsfalse
\includegraphics{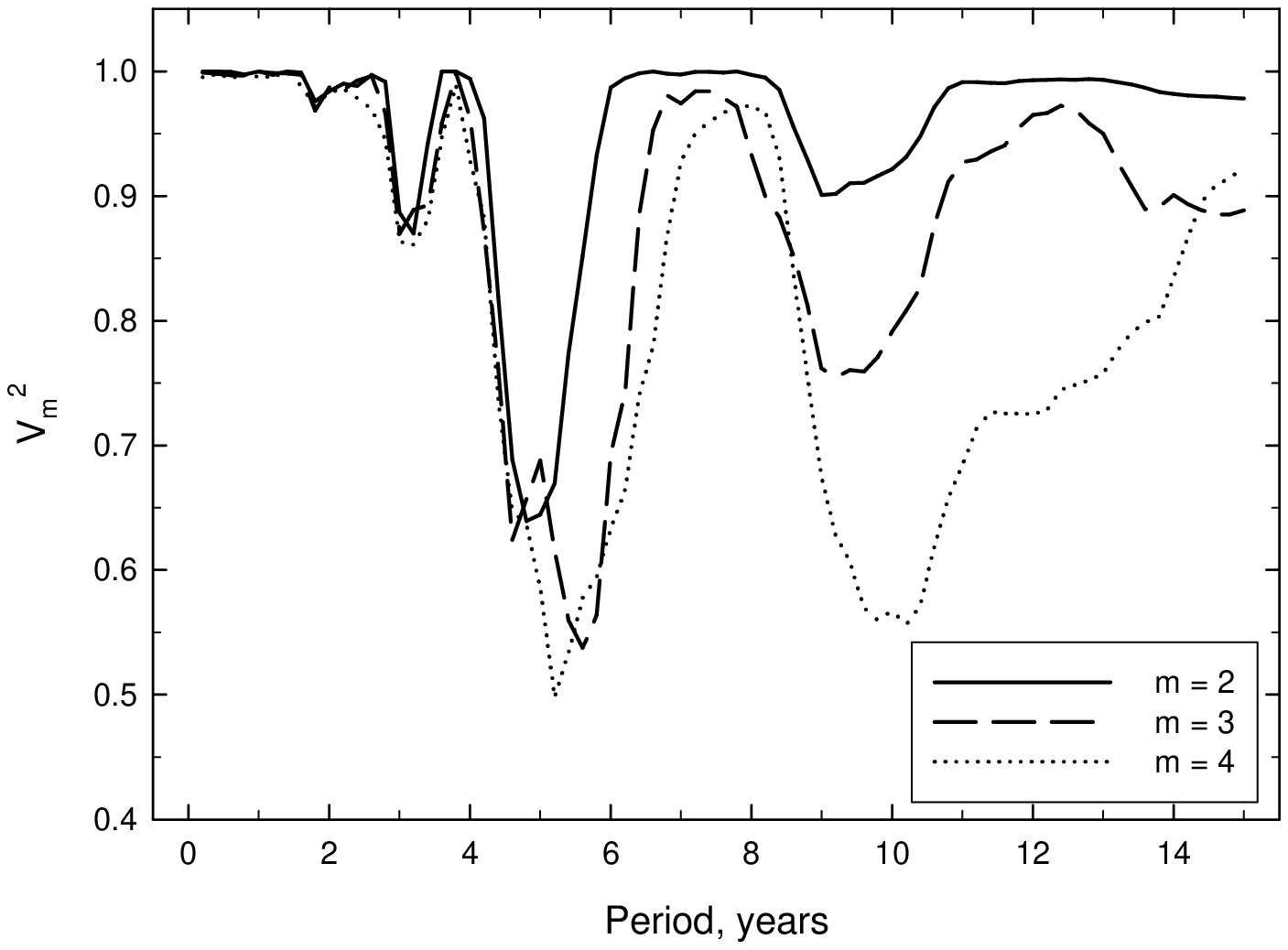}
\captionstyle{normal} \caption{Results of the period search for
2251+158 at 4.8 GHz using the method of Jurkevich. Plots for
various bin numbers for the phase curve are presented.}
\label{2251-jur-4}
\end{figure}

The light curve of 2251+158 is presented in
Fig.~\ref{2251-lcurve}. We can see that powerful flares are
observed approximately every 12 years. Our analysis using the
discrete autocorrelation function detected two periods at all five
frequencies: $6.2 \pm 0.1$ and $12.4 \pm 0.6$ yr
(Table~\ref{table2251}). As an example, Fig.~\ref{2251-dcf-4}
presents the discrete autocorrelation function for the 4.8 GHz
light curve. The method of Jurkevich detected the periods $6.1 \pm
0.6$ and $11.9 \pm 1.1$ yr with high confidence at all frequencies
(Table~\ref{table2251}); this is clearly visible in
Fig.~\ref{2251-jur-4}, which shows a plot of the function
$V_{m}^2$ for the 4.8-GHz light curve.

Thus, the periods $6.2 \pm 0.1$ and $12.4 \pm 0.6$ yr were
detected at all five frequencies by both the method of Jurkevich
and the discrete autocorrelation function. The light curves of
$2251+158$ encompass up to six six-year periods, while only three
12-year periods are included in the most extensive light curve at
8 GHz; further observations are required to verify the presence of
the latter period (Table~\ref{table1}). The period $6.2 \pm 0.1$
yr is in good agreement with the periods reported
earlier~\cite{Ciaramella2004}. Figure~\ref{2251-lcurve} presents
the light curve of $2251+158$ with the positions of the flares
corresponding to the 6.2 and 12.4-year periods marked. The $12.4
\pm 0.6$-year period is determined by the flares of 1967, 1982,
and 1995, while the $6.2 \pm 0.1$-year period is determined by
these flares, together with the somewhat weaker intermediate
flares of 1975 and 1987.

The quasar $2251+158$ is one of the best observed radio sources,
and a large amount of information about its VLBI structure has
been published, making it possible to compare variations in the
source structure with variations in the light curves.
Pauliny-Toth~\cite{pauliny1987,pauliny1998} reported the births of
components at epochs 1966 and $1981.7 \pm 0.8$ at 2.8 cm.
Cawthorne and Gabuzda~\cite{cawthorne1996} reported the probable
birth of a component at epoch 1988.2 based on polarization
observations at 5 GHz. Based on their monitoring of gamma-ray
blazars at 43GHz , Jorstad et al.~\cite{Jorstad2001} derived the
birth epochs for there VLBI components $1994.45 \pm 0.03$ (B1),
$1995.05 \pm 0.05$ (B2), and $1995.59 \pm 0.14$ (B3). The 43-GHz
observations of Gomez et al.~\cite{gomez1999} and Kemball et
al.~\cite{Kemball1996} indicate the possible appearance of
components at epochs 1995.7 and 1994.9, in agreement with the
results of Jorstad et al.~\cite{Jorstad2001}. The arrows in
Fig.~\ref{2251-lcurve} mark the published birth epochs for VLBI
components.We can see that these birth epochs are in good
agreement with the epochs of powerful flares, allowing for the
fact that there were no regular VLBI observations of $2251+158$
before 1981. The presence of several components during the flare
of 1995 can be explained by the good spatial resolution provided
at 43GHz compared to the observations at lower frequencies. Thus,
based on the available data on the birth epochs of VLBI
components, the time interval between the generation of new
components is about 6.3yr , in good agreement with the six-year
period we have found for the radio light curves. The analysis of
the frequency delays and spectral evolution of flares carried out
by Pyatunina et al.~\cite{Pyatunina2004} showed the presence of a
characteristic variability time scale of about 13yr , in agreement
with the detected period of $12.4 \pm 0.6$ yr.

Thus, the periods of $6.2 \pm 0.1$ and $12.4 \pm 0.6$ yr found
formally using the period-search methods are in good agreement
with the characteristic time scale for variations in the VLBI
structure, as well as the frequency delays and spectra of flares.
According to our estimates, the next flare in $2251+158$ should
occur in 2007, have an amplitude comparable to the flare of 1995,
and be accompanied by the appearance of a new VLBI component.
Taking into account the source redshift, $z = 0.859$, the detected
periods in the source rest frame are 3.3 and 6.7 yr.

Note also that our detected period $12.4 \pm 0.6$ yr coincides
with the period of 12.39 yr found by Cheng-yue~\cite{Su2001} for
the hundred-year historical optical light curve in the $B$ band.
The agreement of the periods found in the optical and radio for
$AO~0235+164$~\cite{Raiteri2001} and $BL~Lac$~\cite{villata2004}
may indicate that the same radiation mechanism is responsible for
the variability in these two wavebands. Correlations between the
optical and radio emission of $2251+158$ have been reported by a
number of authors: Pomphrey et al.~\cite{Pomphrey1976} found a
strict correlation with a time delay of 1.2 yr,
Balonek~\cite{Balonek1982}  reported possible correlations with
delays of 180, 285, and 310 days, and Tornikoski et
al.~\cite{Tornikoski1994}  and Hanski et al.~\cite{Hanski2002}
identified simultaneous events in the optical and radio. The
presence of a strict correlation suggests a strict coincidence of
the periods in the two wavebands, but, given the length of the
period, its confirmation requires further observations in both the
optical and radio.

\section{CONCLUSION}
Our analysis of the light curves of $2223-052$, $2230+114$, and
$2251+158$ at 4.8, 8, 14.5, 22, and 37 GHz based on the UMRAO
(USA) and MRAO (Finland)monitoring data has confirmed the presence
of periodicity in the flux variations of these sources. The flux
variations in $2223-052$ occur on characteristic time scales of
$5.8 \pm 0.5$ and $10.9 \pm 0.2$ yr, with the six-year period
being in good agreement with the time scale for the appearance of
VLBI components at 15 GHz. In addition, the detected period of 5.8
yr coincides with the period found earlier for the optical light
curve (5.8 yr)~\cite{barbieri1985}. The light curves for
$2230+114$ show characteristic variation time scales of $4.3 \pm
0.5$ and $8.4 \pm 0.1$ yr, but a comparison with the evolution of
the VLBI structure of this source is complicated by the absence of
observations during the early part of the radio light curves.
Periods of $6.2 \pm 0.1$ and $12.4 \pm 0.6$ yr were detected in
the light curves of $2251+158$, and a comparison with published
VLBI data shows that the 6.2-year period is in good agreement with
the time scale for the evolution of the VLBI structure. The radio
period of $12.4 \pm 0.6$ yr also agrees with the period derived by
Cheng-yue~\cite{Su2001} based on a hundred-year historical optical
light curve.

\section{ACKNOWLEDGMENTS}
This work has made use of the monitoring data of the University of
Michigan Radio Astronomy Observatory. The authors thank Hugh and
Margo Aller for presenting us with these data, as well as for our
long-standing fruitful collaboration. This work has also made us
of monitoring data from the Metsahovi Radio Astronomy Observatory.
We thank Harri Terasranta for fruitful collaboration and providing
us with these data. The work was supported by a grant from the
Ministry of Education of the Russian Federation (grant no. 37840).
N.A. Kudryavtseva thanks the President of the Russian Federation
for financial support in 2004--2005.

\newpage
%
%
%
\newpage
\newpage
%
\newpage
%
%
%
%
%
%
%
%
%
%
\pagebreak
%
\begin{table}[p]
\setcaptionmargin{0mm} \onelinecaptionstrue
\captionstyle{flushleft} \caption{List of objects and
characteristics of the light curves} \label{table1}
\bigskip
\begin{tabular}{|c|c|c|c|c|}
\hline Object & $\nu$, (GHz) & Interval of observations & $\Delta T$, yr & N \\
\hline 2223-052 & 37 & 1986.0 - 2003.6 & 17.6 & 212 \\
(3C 446) & 22 & 1985.0 - 2003.6 & 18.6 & 220 \\
& 14.5 & 1980.0 - 2003.6 & 23.6 & 790 \\
& 8 & 1980.0 - 2003.6 & 23.6 & 735 \\
& 4.8 & 1980.3 - 2003.6 & 23.3 & 520 \\
\hline 2230+114 & 37 & 1986.0 - 2003.2 & 17.2 & 217 \\
(CTA 102) & 22 & 1985.1 - 2003.2 & 18.1 & 288 \\
& 14.5 & 1979.8 - 2003.3 & 23.5 & 725 \\
& 8 & 1974.6 - 2003.2 & 28.6 & 687 \\
& 4.8 & 1981.5 - 2003.1 & 21.6 & 314 \\
\hline 2251+158 & 37 & 1980.7 - 2003.7 & 23.0 & 623 \\
(3C 454.3) & 22 & 1980.5 - 2003.7 & 19.5 & 749 \\
& 14.5 & 1974.1 - 2003.6 & 29.5 & 925 \\
& 8 & 1966.6 - 2003.6 & 37.0 & 1026 \\
& 4.8 & 1978.4 - 2003.6 & 25.2 & 565 \\[1mm]
\hline
\end{tabular}
\end{table}
\newpage
%
\begin{table}[p]
\setcaptionmargin{0mm} \onelinecaptionstrue
\captionstyle{flushleft} \caption{Detected periods for 2223-052
(3T 446)} \label{table2223}
\bigskip
\begin{tabular}{|c|c|c|c|c|}
\hline $\nu$ (GHz) & $P_{Jurk}$,yr & f & $P_{DACF}$,yr & k \\
\hline 37 & $4.2 \pm 0.4$ & 0.26 &  &  \\
& $8.9 \pm 2.5$ & 2.55 & $10.7 \pm 2.5$ & 0.85 \\
\hline
22 & $6.9 \pm 2.2$ & 1.83 &  &  \\
& $11.1 \pm 2.4$ & 2.55 & $10.6 \pm 1.6$ & 0.85 \\
\hline
14.5 & $5.6 \pm 1.3$ & 0.52 &  &  \\
& $9.5 \pm 0.7$ & 1.21 & $11.3 \pm 3.3$ & 0.25 \\
\hline
8 & $5.3 \pm 1.1$ & 0.56 & --- & --- \\
& $9.9 \pm 1.1$ & 0.54 &  &  \\
\hline
4.8 & $7.2 \pm 0.7$ & 0.39 & --- & --- \\[1mm]
\hline
\end{tabular}
\end{table}
\newpage
%
\begin{table}[p]
\setcaptionmargin{0mm} \onelinecaptionstrue
\captionstyle{flushleft} \caption{Detected periods for 2230+114
(CTA~102)} \label{table2230}
\bigskip
\begin{tabular}{|c|c|c|c|c|}
\hline $\nu$ (GHz) & $P_{Jurk}$,yr & f & $P_{DACF}$,yr & k \\
\hline 37 & $5.2 \pm 0.7$ & 0.60 & --- & --- \\
& $9.7 \pm 1.5$ & 0.62 &  &  \\
\hline
22 & $4.9 \pm 0.4$ & 0.60 & --- & --- \\
& $8.4 \pm 2.3$ & 0.47 &  &  \\
\hline
14.5 & &  & $4.3 \pm 0.5$ & 0.41 \\
& $9.2 \pm 2.0$ & 0.22 & $8.3 \pm 1.2$ & 0.17 \\
\hline 4.8 & $3.8 \pm 0.8$ & 0.28 & & \\
& $9.8 \pm 2.1$ & 0.46 & $8.5 \pm 1.2$ & 0.10 \\[1mm]
\hline
\end{tabular}
\end{table}
\newpage
%
\begin{table}[p]
\setcaptionmargin{0mm} \onelinecaptionstrue
\captionstyle{flushleft} \caption{Detected periods for 2251+158
(3C~454.3)} \label{table2251}
\bigskip
\begin{tabular}{|c|c|c|c|c|}
\hline $\nu$ (GHz) & $P_{Jurk}$, yr & f & $P_{DACF}$, yr & k \\
\hline 37 & $5.7 \pm 1.8$ & 0.88 & $6.2 \pm 1.4$ & 0.24 \\
& $11.6 \pm 2.8$ & 1.0 & $12.6 \pm 1.4$ & 0.26 \\
\hline
22 & $6.7 \pm 0.9 $ & 0.35 & $6.3 \pm 1.9$ & 0.14 \\
& $12.1 \pm 0.9 $ & 0.47 & $12.6 \pm 2.4$ & 0.52 \\
\hline
14.5 & $5.7 \pm 1.0$ & 0.86 & $6.2 \pm 2.5$ & 0.24 \\
& $10.5 \pm 0.9$ & 0.45 & $12.5 \pm 1.6$ & 0.63 \\
\hline
8 & $6.7 \pm 0.8$ & 0.19 & $6.1 \pm 3.5$ & 0.15 \\
& $13.5 \pm 1.2$ & 0.44 & $13.4 \pm 1.8$ & 0.45 \\
\hline
4.8 & $5.5 \pm 1.1$ & 0.86 & $6.1 \pm 1.4$ & 0.24 \\
& $11.7 \pm 2.6$ & 0.95 & $11.8 \pm 1.2$ & 0.55 \\[1mm]
\hline
\end{tabular}
\end{table}
\end{document}